\documentclass[
aps%
,groupedaddress%
,nofootinbib%
,showpacs%
,amsfonts%
,nofootinbib%
]{%
revtex4%
}

\usepackage{epsfig}
\usepackage{graphicx}
\usepackage{bm}
\usepackage{amsmath}
\usepackage{amsfonts}
\usepackage{color}
\usepackage{multirow}
\usepackage{undertilde}

\newcommand{\sgn}{{\rm sgn }}
\newcommand{\arctanh}{{\rm arctanh }}

\newcommand{\vnabla}{ \bm \nabla}
\newcommand{\vk}{ \bm k}
\newcommand{\vp}{ \bm p}
\newcommand{\vx}{ \bm x}

\newcommand{\ben}{\begin{enumerate}}
\newcommand{\een}{\end{enumerate}}

\newcommand{\be}{\begin{equation}}
\newcommand{\ee}{\end{equation}}
\newcommand{\bea}{\begin{eqnarray}}
\newcommand{\eea}{\end{eqnarray}}
\newcommand{\ds}{\begin{displaystyle}}
\renewcommand{\ss}{\begin{scriptstyle}}

\newcommand{\Eq}[1]{Eq.~(\ref{eq:#1})}

\newcommand{\ignore}[1]{}

\newcommand{\p}{{\prime}}
\newcommand{\pp}{{\prime\prime}}
\newcommand{\tDelta}{{\tilde{\Delta}}}

\begin{document} 

\title{The perturbative scalar massless propagator in Schwarzschild spacetime}

\author{C. Garc{\'\i}a-Recio}

\author{L. L. Salcedo}

\affiliation{Departamento de F{\'\i}sica At\'omica, Molecular y Nuclear \\
and
  Instituto Carlos I de F{\'\i}sica Te\'orica y Computacional, \\ Universidad
  de Granada, E-18071 Granada, Spain.}

\date{\today}

\begin{abstract}
A short derivation is given of the weak gravitational field approximation to
the scalar massless propagator in Schwarzschild spacetime obtained by Paszko
using the path-integral approach. The contribution from the direct coupling of
the quantum field to the scalar curvature is explictly included.  The
propagator complies with Hadamard's pattern, and the vacuum state is
consistent with the perturbative version of the Boulware vacuum.  The momentum
space propagator is computed for massless or massive particles to the same
perturbative order. The renormalized value of $\langle \phi^2(x)\rangle$ for
the massless case is reproduced.
\end{abstract}



\maketitle

In \cite{Paszko:2012zz}, Paszko carried out a perturbative calculation of the
Feynman propagator of a scalar particle in presence of a Schwarzschild metric,
to first order in the black hole mass, obtaining an explicit analytical
expression for massless particles. A path-integral approach was adopted for
that calculation. In this note we show that this result can be obtained with
considerably less effort by means of the operator formalism.

For simplicity we will work out the Euclidean version and perform the Wick
rotation at the end to obtain the Feynman propagator. We use units $\hbar=c=1$
and $M$, with dimension of length, denotes the black hole mass times Newton's
constant.

Our starting equation is \cite{Birrell:1982ix}
\begin{equation}
(-\square + m^2 + \xi R)_x G(x,x^\p) = \frac{1}{\sqrt{g(x)}}\delta(x-x^\p)
.
\end{equation}
Here $\square=g^{-1/2}\partial_\mu g^{1/2} g^{\mu\nu}\partial_\nu$ is the
Laplacian operator, $R$ the scalar curvature, and $G(x,x^\p)$ is the
(Euclidean) propagator from $x^\p$ to $x$ of the scalar particle with mass
$m$.

Following \cite{Paszko:2012zz}, for the (Euclidean) Schwarzschild metric with
mass $M$, we use isotropic coordinates \cite{Landau:1987gn},
\begin{equation}
g_{\mu\nu} dx^\mu dx^\nu =
\left(\frac{1-\frac{M}{2r}}{1+\frac{M}{2r}}\right)^2 d\tau^2
+\left(1+\frac{M}{2r}\right)^4 d\vx^2 .
\end{equation}
Here we use $\tau$ to denote the Euclidean time (not the proper time), in
order to distinguish it from the Lorentzian time $t$, below.

To first order in an expansion in powers of $M$ the propagator equation
reduces to
\begin{equation}
\left[
-\left(1+\frac{2M}{r}\right)\partial_\tau^2
-\left(1-\frac{2M}{r}\right)\bm{\nabla}^2
+ m^2
+ \xi R
\right]_x G(x,x^\p) = \left(1-\frac{2M}{r}\right) \delta(x-x^\p)
+ O(M^2)
,
\label{eq:3}
\end{equation}
Let us expand also the propagator and the scalar curvature,
\begin{equation}
G = G^{(0)} + G^{(1)} +  O(M^2)
,
\quad
R = R^{(1)} +  O(M^2)
,
\end{equation}
where $G^{(n)}$ and $R^{(n)}$ contain precisely $n$ powers of
$M$. Substituting in \Eq{3} and equating terms with equal powers of $M$
produces the following relations
\begin{eqnarray}
(-\partial_\tau^2 -\bm{\nabla}^2 + m^2 ) G^{(0)}(x,x^\p) &=& \delta(x-x^\p)
,
\\
(-\partial_\tau^2 -\bm{\nabla}^2 + m^2 ) G^{(1)}(x,x^\p) &=&
\left[\frac{4M}{r}(\partial_\tau^2-\frac{1}{2}m^2) - \xi R^{(1)} \right]_x
G^{(0)}(x,x^\p)
.
\end{eqnarray}
From these relations it immediately follows that\footnote{Once the Minkowski
  vacuum has been adopted at zeroth order, the perturbative expansion
  automatically selects a privileged choice of boundary conditions at any
  other perturbative order. We do not override that choice here.}
\begin{equation}
G^{(1)}(x,x^\p) =
\int d^4 x^\pp  G^{(0)}(x,x^\pp)
\left[
\frac{4M}{r^\pp}
(\partial_{\tau^\pp}^2-\frac{1}{2}m^2) 
- \xi R^{(1)}(x^\pp)
\right]
G^{(0)}(x^\pp,x^\p)
.
\label{eq:7}
\end{equation}

\begin{figure}[h]
\begin{center}
\includegraphics[width=0.20\textwidth]{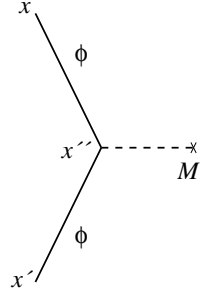}
\caption{\small Feynman graph for \Eq{7}. The vertex can be read off from
  \Eq{8}.}
\label{fig:1}
\end{center}
\end{figure}

Of course, the same result can be obtained by applying standard perturbation
theory to first order. See Fig.\ref{fig:1}. The Feynman rule for the
interaction vertex of the particle with the static gravitational field can be
read off from the Lagrangian:
\begin{eqnarray}
{\mathcal L}(x) &=& \sqrt{g}\left(
\frac{1}{2}g^{\mu\nu}\partial_\mu\phi\partial_\nu\phi 
+ \frac{1}{2}(m^2+\xi R)\phi^2
\right)
\nonumber
\\
&=&
\frac{1}{2}(\partial_\mu\phi)^2+\frac{1}{2}m^2\phi^2
+\frac{1}{2}\frac{4M}{r}\left((\partial_\tau\phi)^2+\frac{1}{2}m^2\phi^2\right)
+ \frac{1}{2}\xi R^{(1)} \phi^2
+O(M^2)
.
\label{eq:8}
\end{eqnarray}
Note that in the Euclidean perturbative computation the metric is just
$\delta_{\mu\nu}$.

In order to obtain a closed expression in $x$-space (as opposed to momentum
space), from now on we consider only the massless case, $m=0$.  In this case
\begin{equation}
G^{(0)}(x,x^\p) = \frac{1}{(2\pi)^2}\frac{1}{(x-x^\p)^2}
.
\end{equation}

As follows from \Eq{7}, the effect of $\xi R$ on the propagator is additive to
first order, so we define
\begin{equation}
G^{(1)}(x,x^\p) = G^{(1)}_0(x,x^\p) + G^{(1)}_\xi(x,x^\p)
.
\end{equation}

Let us consider first the effect of the scalar curvature term. From Einstein's
equations, a Dirac delta mass distribution as source of the Schwarzschild
metric corresponds to a scalar curvature $R(x)=8\pi M \delta(\vx)+O(M^2)$.
When this is inserted in \Eq{7}, one readily obtains
\begin{equation}
G^{(1)}_\xi(x,x^\p) = - \xi \frac{2 M}{(2\pi)^2}\frac{1}{r r^\p} 
\frac{r+r^\p}{\Delta\tau^2+(r+r^\p)^2}
,
\label{eq:11}
\end{equation}
where $\Delta\tau= \tau-\tau^\p$. For the $\xi=0$ part, \Eq{7} reduces to
\begin{equation}
G^{(1)}_0(x,x^\p) =
4M\partial_\tau^2 F(x,x^\p)
,
\end{equation}
with
\begin{equation}
(2\pi)^4 F(x,x^\p)
=
\int d^4 x^\pp  \frac{1}{r^\pp} \frac{1}{(x-x^\pp)^2} \frac{1}{(x^\pp-x^\p)^2}
.
\end{equation}
Here it is appropriate to use a standard Feynman parameterization
\begin{equation}
\frac{1}{ab} = \int_0^1 d\zeta \frac{1}{\left(\zeta a + (1-\zeta) b\right)^2}
,
\end{equation}
which yields
\begin{equation}
(2\pi)^4 F(x,x^\p)
=
\int d^4 x^\pp  \frac{1}{r^\pp}
\int_0^1 d\zeta \frac{1}{((x^\pp-y)^2 + \zeta(1-\zeta) l^2)^2}
,
\end{equation}
where $y= \zeta x+(1-\zeta)x^\p$ and $l^2 = (x-x^\p)^2$. Now it is
straightforward to carry out the $x^\pp$ integration, starting with the
angular part, then $\tau^\pp$ and lastly $r^\pp$. This gives
\begin{equation}
(2\pi)^4 F(x,x^\p)
=
\int_0^1 d\zeta \frac{\pi^2}{|\bm{y}|}
\log\left(\frac
{\sqrt{\bm{y}^2+\zeta(1-\zeta)l^2}+|\bm{y}|}
{\sqrt{\bm{y}^2+\zeta(1-\zeta)l^2}-|\bm{y}|}
\right)
.
\end{equation}
The remaining integral produces a rather complicated expression. However, the
function $F$ itself is not needed. As it turns out, the first time
derivative of $F$ takes a simple form:
\begin{eqnarray}
(4\pi)^2\partial_\tau F(x,x^\p)
&=&
-\frac{2\Delta\tau}{l^2}
\int_0^1 d\zeta 
\frac{1}
{\sqrt{\zeta r^2 + (1-\zeta)r^{\p 2} + \zeta(1-\zeta)\Delta\tau^2}}
\nonumber \\
&=&
-\frac{4}{l^2}
\arctan\left(\frac{\Delta\tau}{r+r^\p}\right)
.
\end{eqnarray}
Taking another time derivative yields the final expression for $G^{(1)}$
(adding the $\xi$ dependent term):
\begin{equation}
G^{(1)}_0(x,x^\p)
=
\frac{1}{(2\pi)^2}\frac{4M}{l^2}
\left(
\frac{2\Delta\tau}{l^2} \arctan\left(\frac{\Delta\tau}{r+r^\p}\right)
-\frac{r+r^\p}{(\Delta\tau)^2+(r+r^\p)^2}
\left(1 + \xi \frac{l^2}{2 r r^\p}\right)
\right)
.
\label{eq:16}
\end{equation}

To obtain the propagator with Lorentzian signature, it only remains to perform
the Wick rotation. This entails taking an analytical continuation
\begin{equation}
i\Delta_F(\Delta t;\vx,\vx^\prime) = 
G(\Delta\tau = i\Delta t;\vx,\vx^\prime)
.
\end{equation}
Specifically, for the Feynman propagator, for positive (negative) $\Delta
t=t-t^\p$, the continuation is to be taken from the positive (negative) ${\rm
  Re}(\Delta\tau)$ half-planes, respectively. Hence, $\Delta\tau = i\Delta t +
\sgn(\Delta t)\eta$, $\eta=0^+$. For the zeroth order, this yields the free
Feynman propagator for a massless scalar particle
\begin{equation}
\Delta^{(0)}_F(x,x^\p) = \frac{i}{(2\pi)^2}\frac{1}{s^2-i\eta},
\end{equation}
where $s^2= \Delta t^2-\Delta\vx^2$ with $\Delta\vx= \vx-\vx^\p$. For the
first order term, the Wick rotation gives (leaving $\eta$ implicit)
\begin{equation}
\Delta_F^{(1)}(x,x^\p)
=
\frac{i}{(2\pi)^2}\frac{4M}{s^2}
\left(
\frac{r+r^\p}{\Delta t^2-(r+r^\p)^2}
\left(1 - \xi \frac{s^2}{2 r r^\p}\right)
+\frac{2\Delta t}{s^2} \arctanh\left(\frac{\Delta t}{r + r^\p}\right)
\right)
.
\label{eq:19}
\end{equation}
This form applies to the case $|\Delta t|<r+r^\p$ (taking for $\arctanh$ the
real branch). When $|\Delta t|>r+r^\p$, the expression is\footnote{This is
  most easily obtained by using first the identity $
  \arctan\left(\frac{1}{x}\right)=-\arctan(x)+\frac{\pi}{2}\sgn(x)$ in
  \Eq{16}, and noting that under Wick rotation $\sgn(\Delta\tau)$ goes to
  $\sgn(\Delta t)$.  }
\begin{equation}
\Delta_F^{(1)}(x,x^\p)
=
\frac{i}{(2\pi)^2}\frac{4M}{s^2}
\left(
\frac{r+r^\p}{\Delta t^2-(r+r^\p)^2}
\left(1 - \xi \frac{s^2}{2 r r^\p}\right)
+\frac{2\Delta t}{s^2} \arctanh\left(\frac{r+r^\p}{\Delta t}\right)
-i\pi\frac{|\Delta t|}{s^2}
\right)
.
\end{equation}
When $\xi=0$, this is the same result already obtained in \cite{Paszko:2012zz}.

We have explictly verified that
$\Delta_F^{(0)}(x,x^\p)+\Delta_F^{(1)}(x,x^\p)$ solves the Klein-Gordon
differential equation to $O(M)$.

From the symmetries of the problem, the propagator will be a function of $r$
and $r^\p$, as well as $s^2$ and $\Delta t$. To $O(M)$, the $\xi=0$ part of
the propagator depends only on the combination $r+r^\p$, but this property is
not preserved by the contribution from $\xi R$.

To the same order in $M$, the explicit result in momentum space can also be
given, even for massive particles, as it follows immediately from the Feynman
vertex in \Eq{8} (in its Lorentzian signature version). Letting
\begin{equation}
D(p)=\frac{1}{p^2-m^2+i\eta},
\end{equation}
the propagator is given by
\begin{eqnarray}
\tDelta_F^{(0)}(p,p^\p) &=& (2\pi)^4\delta(p-p^\p) D(p),
\nonumber 
\\
\tDelta_F^{(1)}(p,p^\p)
 &=&
2\pi\delta(p^0-p^\p{}^0)
8\pi M 
D(p) \left[
\frac{m^2-2p^0p^\p{}^0}{(\vp-\vp^\p)^2}
+ \xi
\right]
D(p^\p).
\label{eq:22}
\end{eqnarray}

Next, we investigate whether the propagator conforms to Hadamard's pattern,
\begin{equation}
-2i(2\pi)^2 \Delta_F(x,x^\p) = \frac{U}{\sigma} + V\log(\sigma) + W
,
\end{equation}
where $U$, $V$ and $W$ are regular functions of $x$ and $x^\p$ at $x=x^\p$ (at
the points where the metric itself is non singular), with $U(x,x)=1$, and
$\sigma(x,x^\p)$ is Synge's function, half the square of the length along the
geodesic joining $x$ and $x^\p$ (and $\sigma \to \sigma-i\eta$ is
implicit). To be a Hadamard state is a well-known requirement for the vacuum
being sufficiently regular in the ultraviolet region to ensure a non singular
stress-energy tensor \cite{Wald:1995yp,Decanini:2005eg}.

The momenta $q=p-p^\p$ and $Q=(p+p^\p)/2$ control, respectively, $(x+x^\p)/2$
and $x-x^\p$ in $x$-space. The large $Q$ behavior $\tDelta_F^{(1)} \sim
Q^0{}^2/Q^4$ competes with that of $\tDelta_F^{(0)} \sim 1/Q^2$, and this
could give rise to a singular coincidence limit in the propagator in
$x$-space. The corresponding situation manifests in
$s^2 \Delta_F^{(1)}(x,x^\p)$, which behaves as $\Delta t^2/s^2$ and so it is not a
continuous function in the coincidence limit.

In order to clarify this issue, $\sigma$ has to be computed to order
$M$. Being an extremal curve, the length of the geodesic can be obtained to
first order by using the zeroth order path, i.e., the straight line. This
gives
\begin{equation}
\sigma(x,x^\p) = \frac{1}{2}s^2 - M (\Delta t^2+\Delta\vx^2)h + O(M^2)
,
\end{equation}
where $h$ is the average value of $1/r$ along the path:
\begin{equation}
h(\vx,\vx^\p) = \int_0^1 \frac{d\lambda}{|\lambda\vx+(1-\lambda)\vx^\p|}
= \frac{2}{|\Delta\vx|} \arctanh \left(\frac{|\Delta\vx|}{r+r^\p}\right)
.
\end{equation}
When combined with the propagator in
\Eq{19}, the following relation obtains
\begin{eqnarray}
-i(2\pi)^2 2\sigma\Delta_F &=& 
1 + 8M\frac{\Delta t^2}{s^2}\left(
\frac{\arctanh \left(\frac{\Delta t}{r+r^\p}\right)}{\Delta t} 
-
\frac{\arctanh \left(\frac{|\Delta\vx|}{r+r^\p}\right)}{|\Delta\vx|} 
\right)
\nonumber\\
&&
+ 4M\left(
\frac{r+r^\p}{\Delta t^2 - (r+r^\p)^2}
\left(1 - \xi \frac{s^2}{2 r r^\p}\right)
 +
\frac{\arctanh \left(\frac{|\Delta\vx|}{r+r^\p}\right)}{|\Delta\vx|} 
\right)
+  O(M^2)
.
\end{eqnarray}
The term with $4M$ is manifestly regular as a function of $\Delta x^\mu$ at
$x=x^\p$.  In the term with $8M$, the structure $(f(\Delta
t^2)-f(\Delta\vx^2))/(\Delta t^2-\Delta\vx^2)$, being $f$ a regular function,
is itself regular. It follows that the propagator is of Hadamard type with
$V=O(M^2)$, and $U(x,x)=1+O(M^2)$ is also checked.

To further expose the vacuum structure, we adopt the point of view that the
Minkowski vacuum, $|0\rangle$, and the corresponding Hamiltonian $H_0$, are
subject to an $O(M)$ time-independent perturbation. Specifically,
$H=H_0+H_1+O(M^2)$ with (the canonical momentum being
$\pi(x)=(1+4M/r)\partial_0\phi(x)+O(M^2)$)
\begin{equation}
H_1 = \int d^3x 
\frac{1}{2} : \left[
\frac{4M}{r} \big( \frac{1}{2} m^2\phi^2(\vx) - \pi^2(\vx) \big)
+ \xi R^{(1)}(\vx) \phi^2(\vx)
\right]:
.
\end{equation}
Straightforward application of time-independent perturbation theory to first
order yields the new perturbed ground state:
\begin{equation}
|\utilde{0}\rangle
=
|0\rangle
-
\int
\frac{d^3k}{(2\pi)^3}\frac{1}{\sqrt{2\omega}}
\frac{d^3k^\p}{(2\pi)^3}\frac{1}{\sqrt{2\omega^\p}}
\frac{8\pi M}{\omega+\omega^\p}
\left[
\frac{\omega\omega^\p+\frac{1}{2}m^2}{|\vk+\vk^\p|^2}
+\frac{\xi}{2}
\right]
a^\dagger(\vk) a^\dagger(\vk^\p)
|0\rangle
+ O(M^2)
,
\end{equation}
where $\omega=\sqrt{\vk^2+m^2}$ and $\omega^\p=\sqrt{\vk^\p{}^2+m^2}$, and
$a^\dagger(\vk)$ is the creation operator of the unperturbed Fock vacuum, with
normalization $[a(\vk),a^\dagger(\vk^\p)]= (2\pi)^3\delta(\vk-\vk^\p)$.

The perturbed vacuum is annihilated by the perturbed annihilation operators
\begin{equation}
\tilde{a}(\vk) = a(\vk) + 
\int \frac{d^3k^\p}{(2\pi)^3}
\frac{1}{\sqrt{\omega\omega^\p}}
\left(
\frac{8\pi M}{\omega - \omega^\p}
\left[
\frac{\frac{1}{2}m^2-\omega\omega^\p}{|\vk-\vk^\p|^2}
+\frac{\xi}{2}
\right]
a(\vk^\prime)
+
\frac{8\pi M}{\omega + \omega^\p}
\left[
\frac{\frac{1}{2}m^2+\omega\omega^\p}{|\vk+\vk^\p|^2}
+\frac{\xi}{2}
\right]
a^\dagger(\vk^\prime)
\right)
+ O(M^2).
\end{equation}
These operators satisfy $[H,\tilde{a}(\vk)] = -\omega \tilde{a}(\vk)$ to
$O(M)$. The single particle spectrum is unchanged since it is determined by
the asymptotic scattering region, far from the black hole. The fact that
$\tilde{a}(\vk)|\utilde{0}\rangle = 0$ and the commutator of the perturbed
creation and annihilation operators are still c-numbers allows to use Wick's
theorem for the field correlation functions, and this implies that the state
$|\utilde{0}\rangle$ is Gaussian.

A direct calculation shows that the Fourier transform of
$-i\langle\utilde{0}|T(\phi(x)\phi(x^\p)) |\utilde{0}\rangle$ just reproduces
the propagator in \Eq{22}. (Note that $H$, and not $H_0$, has to be used here
to evolve the fields in Heisenberg image.) The previous arguments strongly
suggest that $|\utilde{0}\rangle$ is just the Boulware vacuum, the stationary
non thermal and non radiating ground state of the Schwarzschild metric
\cite{Boulware:1974dm,Christensen:1977jc}, albeit to $O(M)$. Therefore, the
propagator applies to a static spherical star, in the regime $r \gg M$, rather
than to a proper black hole, which would be in a thermal state with a
temperature $T=1/(8\pi M)$ \cite{Hartle:1976tp}.

The renormalized value of $\langle \phi^2(x)\rangle$ has been computed in
\cite{Satz:2004hf} for a massless particle in the weak gravitational field of
a static spherical star. Local approximations have been avoided in
\cite{Satz:2004hf}.

From the definition of the propagator, the bare value of $\langle
\phi^2(x)\rangle$ follows from taking the coincidence limit in $G(x,x^\p)$ or
$i\Delta_F(x,x^\p)$. Here we take the alternative approach of going back to
\Eq{7} and setting $x=x^\p$ there (for $m=0$). Straightforward integration
over $\tau^\pp$ gives
\begin{equation}
G^{(1)}(x,x) = 
-\frac{\pi M}{(2\pi)^4}
\int d^3 x^\pp \left[
\frac{1}{|\vx^\pp|}\frac{1}{|\vx-\vx^\pp|^5}
+
4\pi\xi \delta(\vx^\pp) \frac{1}{|\vx-\vx^\pp|^3}
\right]
.
\end{equation}
The term with $\xi$ is ultraviolet finite and it just reproduces the result of
taking the coincidence limit directly in \Eq{11}. The term without $\xi$ is
divergent. However, as a distribution $|\vx-\vx^\pp|^{-5} =
(1/6)\vnabla^2|\vx-\vx^\pp|^{-3}$. Integration by parts and use of
$\vnabla^2|\vx^\pp|^{-1} = -4\pi\delta(\vx^\pp)$ immediately gives
\begin{equation}
\langle \phi^2(x)\rangle
=
- \left(\xi-\frac{1}{6}\right) \frac{1}{(2\pi)^2} \frac{M}{r^3}
+ O(M^2)
.
\label{eq:33}
\end{equation}
This result is in agreement with that obtained in \cite{Satz:2004hf} by using
dimensional regularization plus standard subtraction of the singular part
\cite{Birrell:1982ix}.

The Feynman propagator at $\vx=\vx^\p$ but arbitrary $\Delta t$, for the same
setting (weak static spherical gravitational field and massless quantum field)
follows from the results in \cite{Louko:2007mu}. Use of $R^{(1)}(x) = 8\pi
M\delta(\vx)$ in Eq.~(6.9) of \cite{Louko:2007mu}, yields a result that almost
agrees with our value for $\Delta_F^{(1)}(t,\vx;t^\p,\vx)$ from
\Eq{19}.\footnote{Both results would agree if a factor 1/2 were added to the
  middle term in the integrand of Eq.~(6.9). It can be noted that this middle
  term would correspond to a Dirac delta function in Eq.~(6.8) of
  \cite{Louko:2007mu}, rather than to an ordinary function.}  The result in
\cite{Louko:2007mu} is non singular in the limit $\Delta t\to 0$, and it is
consistent with \Eq{33}. Instead, we obtain the same non singular part plus a
divergent contribution:
\begin{equation}
 i \Delta_F^{(1)}(t,\vx;t^\p,\vx)\big|_{{\rm div}} =
- \frac{1}{(2\pi)^2} \frac{2M}{r} \frac{1}{\Delta t^2}
.
\end{equation}
This expression is consistent with Eq.~(A1) of \cite{Breen:2011aa}.

As a final comment, we note that ultraviolet divergences are expected to arise
at higher orders in a strict expansion of the propagator in powers of
$M$. These would come from small radii in the intermediate point integrations.
Inspection of the integrals involved suggests that such an expansion breaks
down already at $O(M^3)$ due to the presence of terms
$(M/r^{\pp})^3$.
\ignore{\footnote{Painlev\'e-Gullstrand coordinates
  \cite{Martel:2000rn} would produce an expansion of the propagator in powers
  of $M^{1/2}$. Since the order $M^2$ in isotropic coordinates seems to be
  finite, the first genuine (i.e., not removable by a change of coordinates)
  odd power in $M^{1/2}$ it the propagator can arise at $O(M^{5/2})$. Such
  hypothetical genuine term would certainly translate into a divergence in the
  $M^3$ coefficient if one insisted in an expansion using only integer powers
  of $M$. A divergence at $O(M^3)$ could also be of physical origin and so
  independent of the coordinates used to express the propagator.}
}

\begin{acknowledgments}
This work has been supported by Spanish DGI and FEDER funds (grant
FIS2011-24149), Junta de Andaluc{\'\i}a (grant FQM-225), the Spanish
Consolider-Ingenio 2010 Programme CPAN (CSD2007-00042), and by the EU
HadronPhysics2 project, (grant 227431). The authors benefited from exchange of
ideas by the ESF Research Network CASIMIR.
\end{acknowledgments}

\bibliographystyle{h-physrev4}

\end{document}